\newif\ifDraft\Draftfalse
\newif\ifAnon\Anonfalse
\newif\ifNotes\Notestrue

\documentclass[runningheads,a4paper]{llncs}
\usepackage{tugraz_defaults}
\usepackage{colortbl}
\usepackage{tablefootnote}
\usepackage{multicol}

\title{Remote Scheduler Contention Attacks (Extended Version)}

\ifAnon
\author{}
\institute{}
\else
\author{Stefan Gast \and Jonas Juffinger \and Lukas Maar \and Christoph Royer \and Andreas Kogler \and Daniel Gruss}
\institute{Graz University of Technology, Austria}
\authorrunning{S. Gast \and J. Juffinger \and L. Maar \and C. Royer \and A. Kogler \and D. Gruss}
\fi

\newcommand{\tidle}{t_{\mathrm{idle}}}
\newcommand{\smin}{n_{\mathrm{min}}}

\newcommand{\highresapi}{High Resolution Time API\xspace}

\newcommand{\CovertErrorZenThreeValue}{1.46}
\newcommand{\CovertErrorZenThreeUnit}{\percent}

\newcommand{\CovertErrorZenThree}{\SIs{\CovertErrorZenThreeValue}{15}{0.14}{\CovertErrorZenThreeUnit}}

\newcommand{\CovertTrueCapacityZenThreeValue}{891.9}
\newcommand{\CovertTrueCapacityZenThreeUnit}{\bit\per\second}
\newcommand{\CovertTrueCapacityZenThreeShort}{\SI{\CovertTrueCapacityZenThreeValue}{\CovertTrueCapacityZenThreeUnit}}
\newcommand{\CovertTrueCapacityZenThree}{\SIs{\CovertTrueCapacityZenThreeValue}{15}{8.75}{\CovertTrueCapacityZenThreeUnit}}

\newcommand{\CovertErrorZenFourValue}{0.69}
\newcommand{\CovertErrorZenFourUnit}{\percent}
\newcommand{\CovertErrorZenFourShort}{\SI{\CovertErrorZenFourValue}{\CovertErrorZenFourUnit}}
\newcommand{\CovertErrorZenFour}{\SIs{\CovertErrorZenFourValue}{10}{0.08}{\CovertErrorZenFourUnit}}

\newcommand{\CovertTrueCapacityZenFourValue}{940.7}
\newcommand{\CovertTrueCapacityZenFourUnit}{\bit\per\second}
\newcommand{\CovertTrueCapacityZenFourShort}{\SI{\CovertTrueCapacityZenFourValue}{\CovertTrueCapacityZenFourUnit}}
\newcommand{\CovertTrueCapacityZenFour}{\SIs{\CovertTrueCapacityZenFourValue}{10}{5.15}{\CovertTrueCapacityZenFourUnit}}

\colorlet{acolor}{blue!20}
\colorlet{bcolor}{green!20}

\definecolor{brass}{rgb}{0.722, 0.584, 0.102}
\definecolor{ocean}{rgb}{0.137, 0.722, 0.627}
\definecolor{wine}{rgb}{0.941, 0.318, 0.329}

\newcolumntype{C}[1]{>{\raggedleft\arraybackslash}p{#1}}

\begin{document}

\maketitle

\begin{abstract}
In this paper, we investigate unexplored aspects of scheduler contention:
We systematically study the leakage of all scheduler queues on AMD Zen 3 and show that all queues leak.
We mount the first scheduler contention attacks on Zen 4, with a novel measurement method evoking an out-of-order race condition, more precise than the state of the art.
We demonstrate the first inter-keystroke timing attacks based on scheduler contention, with an F$_1$ score of \SI{\geq 99.5}{\percent} and a standard deviation below \SI{4}{\milli\second} from the ground truth.
Our end-to-end JavaScript attack transmits across Firefox instances, bypassing cross-origin policies and site isolation, with \CovertTrueCapacityZenThreeShort{} (Zen 3) and \CovertTrueCapacityZenFourShort{} (Zen 4).
\end{abstract}

\section{Introduction}
Modern CPUs execute micro-operations~(\uops) out-of-order to improve performance.
To select which \uops to execute next, modern CPUs have one or more schedulers~\cite{Johnson2021M1Firestorm,AMD2020OptimizationEPYC7002,AMD2020OptimizationEPYC7003,AMD2023OptimizationZen4}.
Gast~\etal\cite{Gast2023SQUIP} showed that an attacker-controlled binary, executed natively on the machine, can exploit the integer multiplication scheduler queue on AMD CPUs to leak cryptographic keys.

Side-channel attacks mounted via the browser are considered more dangerous, as a victim only has to visit a malicious website~\cite{Oren2015}.
However, it is unclear if scheduler contention attacks are possible from JavaScript.
The shown native code attacks focus on highly repetitive events, such as covert channels and encryptions.
This raises the question whether singular low frequency events can also be monitored using scheduler contention.
Additionally, the attack surface of other schedulers than the integer multiplication scheduler is unexplored.
Furthermore, although AMD acknowledged the security issues~\cite{AMD2022Execution} reported by Gast~\etal\cite{Gast2023SQUIP}, it is unclear whether AMD Zen 4 is also affected.

In this paper, we systematically study scheduler-contention side channels in both native and JavaScript contexts, using AMD's Zen 3 and Zen 4 architectures:
\begin{compactitem}
  \item[\textbf{RQ1}:] What distinguishes schedulers, \eg priming instructions or queue sizes?
  \item[\textbf{RQ2}:] Which measurement methods are practical on recent processors?
  \item[\textbf{RQ3}:] Which scheduler queues leak non-repeatable events, \eg keystrokes?
  \item[\textbf{RQ4}:] Can a remote attacker exploit scheduler contention using JavaScript?
\end{compactitem}
We systematically address these overarching research questions:

For \textbf{RQ1}, we determine generic requirements to prime arbitrary schedulers, resulting in effective priming sequences for all execution units on Zen 3.
On Zen 4, we found entirely reimplemented instructions, diverging from the behavior on Zen 3 and, based on reverse-engineering that contradicts AMD's documentation.

For \textbf{RQ2}, we evaluate bingo race, a timingless, more precise, out-of-order race condition based method to measure scheduler contention.
This method yields the correct scheduler~1 capacity on both Zen 3 and Zen 4, in contrast to prior work.

For \textbf{RQ3}, we present keystroke-timing attacks on all integer scheduler queues.
We leak password keystrokes on a login screen with an F$_1$ score of \SI{\geq 99.5}{\percent}.
For the correctly detected keystrokes, the standard deviation from the ground truth is below \SI{4}{\milli\second}, confirming the high precision of our attack.

Finally, for \textbf{RQ4}, we present a \SI{1}{\kilo\bit\per\second} JavaScript scheduler contention covert channel in Firefox with an error rate of less than \SI{1.5}{\percent} (Zen 3) and \SI{0.7}{\percent} (Zen 4).
This results in a true capacity of \CovertTrueCapacityZenThreeShort{} (Zen 3) and \CovertTrueCapacityZenFourShort{} (Zen 4).
Our attack works across websites, with sender and receiver running in different browser windows, bypassing cross-origin policies and site isolation.

\begin{table}[t]
\centering
    \caption{Comparison of browser-based covert channels}
    \label{tab:covert-compare}
    \def\tabcolsep{4.5pt}%
    \scriptsize
    \begin{tabular}{lrrrc}
      \toprule
        \textbf{Covert Channel} & \textbf{Raw Capacity}\footnotemark[1] & \textbf{Error Rate} & \textbf{True Capacity} &  \\
      \midrule
        Prime+Probe~\cite{Oren2015}\footnotemark[2] & \SI{320}{\kilo\bit\per\second} & - & - & \xmark \\
        \textbf{Our work} & \SI{1000}{\bit\per\second} & \CovertErrorZenFourShort & \CovertTrueCapacityZenFourShort & \cmark \\
        Port contention~\cite{Rokicki2022webport} & \SI{200}{\bit\per\second} & \SI{1}{\percent} & \SI{184}{\bit\per\second} & \cmark \\
        Event loop~\cite{Vila2017} & \SI{200}{\bit\per\second} & - & - & \xmark \\
        Hardware interrupts~\cite{Lipp2017practical} & \SI{25}{\bit\per\second} & - & - & \cmark \\
        DRAM~\cite{Schwarz2017Timers} & \SI{11}{\bit\per\second} & \SI{0}{\percent} & \SI{11}{\bit\per\second} & \cmark \\
        RIDL (Evict+Reload)~\cite{VanSchaik2019RIDL} & \SI{8}{\bit\per\second} & - & - & \xmark \\
        Disk contention~\cite{VanGoethem2017isolated} & \SI{0.5}{\bit\per\second} & \SI{0}{\percent} & \SI{0.5}{\bit\per\second} & \cmark \\
        Memory throttling~\cite{Rushanan2016malloryworker} & \SI{0.2}{\bit\per\second} & \SI{0}{\percent} & \SI{0.2}{\bit\per\second} & \cmark \\
      \bottomrule
    \end{tabular}
    \vspace{0.25cm}
  \begin{minipage}{0.85\hsize}
    \scriptsize{The last column indicates an evaluation with current mitigations. \\
    \textsuperscript{1}\,Sorted by the bandwidth each work reported (\cf~\cite{Rokicki2022webport}). \\
    \textsuperscript{2}\,The work predates current mitigations, including heavy countermeasures against timing attacks. If reimplemented it will likely yield a much lower bandwidth.}
  \end{minipage}
\end{table}
  
In summary, we make the following contributions:
\begin{compactenum}
\item We systematically analyze contention on each scheduler on Zen 3 and 4, revealing effective priming sequences for all of them.
\item We present bingo race, a novel timingless and more accurate measurement method based on out-of-order memory reads.
\item We show that contention attacks on each integer scheduler can observe singular events like inter-keystroke timings of a password entry, with F$_1$ scores \SI{\geq 99.5}{\percent} and a standard deviation below \SI{4}{\milli\second} from the ground-truth timings.
\item We present a scheduler contention covert channel purely in JavaScript running in Firefox 114, with a true capacity of \CovertTrueCapacityZenFour, bypassing cross-origin policies and site isolation.
\end{compactenum}

\paragrabf{Responsible Disclosure}
We reported our findings to AMD on August 8th, 2023.
They acknowledged our findings on August 16th, 2023.

\paragrabf{Outline}
We present background in \Cref{sec:background}, a systematic scheduler queue analysis in~\Cref{sec:queues}, and our bingo race measurement in~\Cref{sec:measurement}.
We present a keystroke timing attack in~\Cref{sec:keystrokes}, a JavaScript covert channel in~\Cref{sec:covert}, and discuss mitigations and related work in~\Cref{sec:discussion}.
\Cref{sec:conclusion} concludes.

\section{Background}\label{sec:background}

In this section, we provide background on out-of-order execution, scheduler contention attacks, and timing measurements in JavaScript.

\paragrabf{Superscalar CPUs}
Superscalar CPUs increase the number of instructions per clock cycle through parallel processing~\cite{Fog2021microarchitecture,AMD2023OptimizationZen4}.
The CPU frontend decodes fetched instructions into micro-ops (\uops) and stores them in the retire control unit (RCU).
The backend has multiple execution units, such as arithmetic and logic units (ALUs), branch execution units (BRUs) and address generation units (AGUs).
Execution units have a scheduler, tracking operand dependencies to determine \uops ready for execution.
Finally, the out-of-order executed \uops retire in instruction stream order, making their results visible in the architectural state.

\paragrabf{Execution Unit Schedulers}\label{sub:schedulers}
A CPU can have one scheduler, such as Intel CPUs~\cite{Intel_opt}, or multiple schedulers, such as AMD~\cite{AMD2020OptimizationEPYC7002,AMD2020OptimizationEPYC7003} and Apple CPUs~\cite{Johnson2021M1Firestorm}.
AMD Zen 2 CPUs have separate schedulers for each ALU and a dedicated scheduler for all AGUs~\cite{AMD2020OptimizationEPYC7002}.
Zen 3~\cite{AMD2020OptimizationEPYC7003} and 4~\cite{AMD2023OptimizationZen4} have schedulers for pairs of ALU and AGU, or ALU and BRU.
Each scheduler has a queue buffering \uops until the execution unit and the input operands are ready.
On Zen 3, the integer execution unit schedulers have a capacity of 24 \uops~\cite{AMD2020OptimizationEPYC7003,Gast2023SQUIP}.
Scheduler contention refers to the situation when a \uop is about to be enqueued into an already full queue~\cite{Gast2023SQUIP}.

\paragrabf{Simultaneous Multithreading (SMT)}\label{sub:smt}
Efficiency is maximized if all execution units are in use.
This is usually not the case with a single instruction stream.
Therefore, many modern CPUs execute multiple instruction streams (2 on AMD Zen CPUs) simultaneously on the same core, sharing the L1 and L2 cache, execution units and schedulers.
This sharing enables various side channels~\cite{Szefer2019survey}.
Scheduler contention attacks exploit the shared scheduler queues~\cite{Gast2023SQUIP}.
Different hardware resources are partitioned in different ways: competitively, each thread can fully use a resource; watermarking, a small fraction is reserved for each thread; static partitioning, each thread can only use their fixed fraction.

\paragrabf{Scheduler Contention Attacks}
Gast~\etal\cite{Gast2023SQUIP} have shown that contention on scheduler queues measurably delays program execution.
A full or almost full queue causes a back-end stall, delaying subsequent \uops.
Their attack on Zen 2 and 3 exploited that unprivileged \texttt{rdpru} timer reads are executed out-of-order, unless the back-end stalls.
As the schedulers are shared between SMT threads, it is possible to observe multiplications of a co-located program.
They demonstrate a covert channel and an attack on square-and-multiply RSA, both in native code.
They also show how to fill (\ie prime) the scheduler used for divisions.
However, they did not investigate all scheduler queues nor JavaScript attacks.

\paragrabf{JavaScript}\label{sec:bg_js}
JavaScript is a just-in-time compiled scripting language for the web that operates within a strict sandbox.
JavaScript code cannot access high-resolution timers via \texttt{rdtsc} or similar instructions.
It instead relies on the \highresapi~\cite{W3C2016}.
To address security concerns associated with timing attacks~\cite{Oren2015,MozillaBug2015}, all major browsers reduced the frequency to \SI{200}{\kilo\hertz}~\cite{MozillaBug2015,ChromiumBug2015,WebkitBug2015}.
However, side channel attacks were still demonstrated~\cite{Kohlbrenner2016,Schwarz2017Timers,Rokicki2022webport,Purnal2023showtime}.
JavaScript supports multithreading through web workers~\cite{Mozilla2023WebWorker}, typically using shared array buffers, representing shared memory~\cite{Mozilla2022Shared}, for communication between threads.

\section{Systematic Analysis of Zen 3 and 4 Scheduler Queues}\label{sec:queues}
Prior work~\cite{Gast2023SQUIP} showed that scheduler queues 0 and 1 can be primed with divisions and multiplications, respectively.
To answer \textbf{RQ1}, we explore priming instructions for integer execution-unit schedulers 2 and 3, as well as the floating point schedulers.
Thus, our analysis fills the gap to a complete coverage of all schedulers on Zen 3 and 4.
Zen 4 has a similar scheduler design to Zen 3, also regarding capacities and execution unit connections~\cite{AMD2020OptimizationEPYC7003,AMD2023OptimizationZen4}.
Thus, we expect similar scheduler queue sizes and usages for Zen 3 and Zen 4.

For each scheduler queue, we need to find a \emph{priming} instruction suitable to fill it.
This instruction must be \emph{delayable}, \emph{targeted}, \emph{single-queue}, \emph{non-serializing}, \emph{unprivileged}, and preferably \emph{single-\uop{}}:
The instruction has to be \emph{delayable} by a long-latency operation to maintain the desired queue occupancy level.
To target a specific scheduler, the instruction must be decoded to a \uop{} that can only be executed on one specific execution unit (\emph{targeted}).
Additionally, the instruction must not cause contention on other queues (\emph{single-queue}).
Furthermore, the instruction must be \emph{non-serializing}, as observing scheduler contention relies on out-of-order execution.
The instruction must be \emph{unprivileged} to allow exploitation from user space.
Finally, for precise control over the occupancy level of the target queue, the instruction must not have more than one input-dependent \uop{} for the targeted scheduler (\emph{single-\uop{}}).
If decoded into multiple \uops{}, the other \uops{} must either be independent or executable by more than one execution unit.

We compose a set of candidate instructions from a complete x86 instruction table by eliminating instructions that do not fulfill all requirements in two phases:

In the \textbf{first} phase, we select a set of candidate instructions that are \emph{delayable}, \emph{non-serializing}, and \emph{unprivileged} from AMD's instruction latency table~\cite{AMD2020OptimizationEPYC7003} and uops.info~\cite{Abel2019uops}.
We initially focus on instructions that are explicitly documented to require the execution unit that is connected to the targeted scheduler.
However, for many instructions the used execution unit is undocumented, \eg \texttt{bsf}, \texttt{bsr}, and the majority of microcoded instructions.
Furthermore, as shown later, there are some discrepancies between the documented execution unit usage and our measurement results, indicating inaccurate or outdated documentation.
Therefore, if we do not find a suitable instruction in the first phase, we extend our search to instructions without documentation about the execution units used.
We start with instructions that are only decoded into a few \uops, as we expect these to allow us to control the occupancy level more precisely (\emph{single-\uop{}}).

In the \textbf{second} phase, we check each candidate instruction whether it is \emph{targeted}, \emph{single-\uop{}}, and \emph{single-queue}, by replacing the multiplications in~\cref{fig:squip-bingo} with a varying number of repetitions of the chosen instruction.
We monitor which scheduler queues are affected by contention via the performance counters \texttt{IntSch[0-3]TokenStall} and \texttt{FPSchRsrcStall}.
To obtain the exact number of repetitions $k$ required to fill the target queue, we introduce a precise contention measurement approach using an out-of-order read from a bingo variable in~\cref{sub:bingothread}.
If we observe $k$ to be more than the capacity (24 \uops~\cite{AMD2020WhereGamingBegins}) of a single scheduler queue, we conclude the instruction is not \emph{targeted}.
If we observe $k$ to be significantly less than the capacity of a single scheduler queue, we conclude the instruction is not \emph{single-\uop{}}.
If we observe multiple performance counters to be increasing with an increased $k$, we conclude the instruction is not \emph{single-queue}.

For the sake of brevity, we do not go into detail for integer scheduler queues 0 and 1, as instruction sequences for these were already documented by Gast~\etal\cite{Gast2023SQUIP}.
In the following, we discuss the results for the remaining scheduler queues.

\subsection{Scheduler Queue 2}\label{sub:sched2}
While AMD's instruction latency table~\cite{AMD2020OptimizationEPYC7003} lists several instructions that are executed on ALU2, we found the microcoded \texttt{stosb} to be the best candidate.

Instructions moving data from a general purpose to a floating point register (\eg \texttt{movd}, \texttt{vmovd} and \texttt{cvtsi2sd}) are documented to be executed on ALU2.
However, contrary to the documentation, our measurements clearly show that they use ALU0 instead.
The instruction latency table also states that bit shift and bit rotation operations, \eg \texttt{rol} or \texttt{shl}, can be scheduled on ALU2 and ALU1, violating our \emph{single-queue} requirement, as we also experimentally confirmed.

We tested microcoded instructions and found \texttt{stosb}, with its implicit \texttt{al} input operand, to cause contention on scheduler 2.
We measure~\cite{Abel2020nanobench} that \texttt{stosb} is decoded into 3 \uops:
One \uop is always enqueued into scheduler queue 2, but not the other 2 \uops.
We observe $k$ to be 22 (see \cref{sec:measurement} for why it is not the documented 24), showing that we can precisely prime scheduler queue 2.

\subsection{Scheduler Queue 3}\label{sub:sched3}
The instruction latency table does not report any instructions that exclusively use ALU3.
Our search again resulted in a microcoded instruction, \texttt{lodsb}, which loads \texttt{[rsi]} into the \texttt{al} register, then increments \texttt{rsi}.
The \texttt{loadsb} instruction is \emph{delayable}, as it performs a partial register load.
When a byte is loaded into the \texttt{al} register, \ie bits 0--7 of the \texttt{rax} register, the remaining bits of \texttt{rax} retain their previous values, creating a dependency on \texttt{rax}~\cite{Fog2021microarchitecture}.
According to nanoBench~\cite{Abel2020nanobench}, \texttt{lodsb} is decoded into 4 \uops.
Our measurements show that one \uop is always enqueued into scheduler 3 while the other 3 \uops are scheduled to other execution units.
$k$ is again 22, showing that we can also precisely prime scheduler 3.

We obtain identical results for \texttt{lodsw}, \texttt{lodsd} and \texttt{lodsq}, loading 16, 32 and 64 bit into \texttt{ax}, \texttt{eax} and \texttt{rax}, respectively.
This is surprising for the 32 and 64 bit variants, \texttt{lodsd} and \texttt{lodsq}, as they completely overwrite \texttt{rax}.
Thus, these instructions likely have a false dependency.

We also evaluated the non-microcoded \texttt{bsf} and \texttt{bsr} instructions that we found to be executed on ALU3 on Zen~3.
Our measurements show that they indeed occupy scheduler 3, however $k$ is only 7.
With a scheduler capacity of 24 \uops~\cite{AMD2020WhereGamingBegins}, this indicates that each of these instructions generates 3 \uops for ALU3, violating our \emph{single-\uop{}} requirement.
In contrast, on Zen~4 the two instructions appear to be decoded into only a single \uop{} that can be executed on every of the four integer ALUs, violating the \emph{targeted} requirement.

\subsection{Observing Scheduler Contention on the FPU}\label{sub:fpu}
In addition to the integer execution unit schedulers, the floating point unit (FPU) schedulers of Zen~3 and 4 can also be primed.
The FPU has 2 schedulers with 32 entries each~\cite{AMD2020OptimizationEPYC7003,AMD2023OptimizationZen4}.
Each scheduler is connected to 3 execution units.
In contrast to the integer execution unit, the FPU has an additional 64 \uops{} non-scheduling queue, buffering \uops between the RCU and the FPU schedulers~\cite{AMD2020OptimizationEPYC7003}.
Hence, we expect to observe stalls when exceeding $64+2\cdot32=128$ repetitions of a priming instruction that can be enqueued into both schedulers.

When priming the FPU schedulers, we can no longer use \texttt{sqrtsd} for the delaying dependency chain, as it is executed on the FPU and influences the measurement.
We move it to the integer unit, by chaining 18 \texttt{div} instructions and converting the final result to floating point, using \texttt{cvtsi2sd}.
According to AMD~\cite{AMD2020OptimizationEPYC7003}, \texttt{cvtsi2sd} is decoded into 2 \uops with one of them utilizing the FPU, reducing the observable capacity by one.
We verify that we can prime both FPU scheduler queues and the non-scheduling queue, using \texttt{vaddsd} as the priming instruction, which targets both FPU schedulers.
With this, we observe back-end stalls with $k>127$, showing that we can fill the queues exactly to their capacity.

To target a single FPU scheduler, we search for an instruction that can only be executed by a single execution unit.
According to AMD~\cite{AMD2020OptimizationEPYC7003}, almost all operations can be handled by 2 or 4 execution units, which are evenly distributed across the schedulers.
The only documented exceptions are \texttt{divsd} and \texttt{sqrtsd}.
Hence, we expect to observe a lower $k$ but again measure a capacity of $k=127$.
This indicates that these operations are executed on multiple execution units, connected to both schedulers, contrary to the documentation.
A possible explanation is that both schedulers have a uniform set of FPU execution units.
Therefore, each FPU scheduler can handle all FPU \uops, making priming a single scheduler impossible.
As demonstrated, it is however possible to prime and probe both schedulers together.

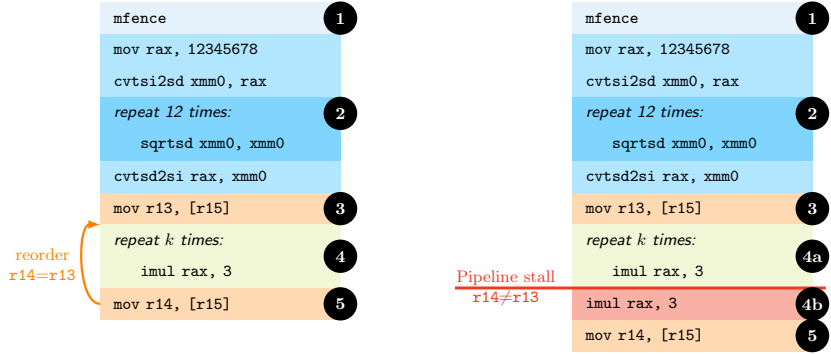
\begin{figure}[t]
  \centering
  \begin{subfigure}[t]{0.45\hsize}
    \resizebox{\hsize}{!}{
      \begin{tikzpicture}
\definecolor{blue1}{HTML}{00aaff}
\definecolor{blue2}{HTML}{007fd0}
\colorlet{drain_bg}{blue2!10}
\colorlet{delay_bg}{blue1!30}
\colorlet{delay_rept_bg}{blue1!50}
\colorlet{read_bg}{orange!30}
\colorlet{fill_free}{green!20}
\tikzset{
    codeblock/.style={
        anchor=north,
        align=left,
        font=\ttfamily,
        minimum width=4.5cm,
        minimum height=0.6cm,
        text width=4cm,
        line width=0pt,
        inner ysep=0pt
    },
    reorder_arrow/.style={
        -latex,
        very thick,
        orange
    },
    reorder_caption/.style={
        midway,
        align=center,
        left
    },
    block_label/.style={
        circle,
        draw=none,
        fill=black,
        text=white,
        font=\bfseries\footnotesize,
        inner sep=0.3mm,
        minimum width=6.3mm
    }
}

\path (-4.5,0.25) rectangle (3,-6.75) ;

\node[codeblock,fill=drain_bg] at (0,0) (mfence) {mfence} ;
\node[codeblock,fill=delay_bg,below=0pt of mfence] (mov_rax) {mov rax, 12345678} ;
\node[codeblock,fill=delay_bg,below=0pt of mov_rax] (cvtsi2sd) {cvtsi2sd xmm0, rax} ;
\node[codeblock,fill=delay_rept_bg,font=\sffamily\itshape,below=0pt of cvtsi2sd] (rept_sqrt) {repeat 12 times:} ;
\node[codeblock,fill=delay_rept_bg,below=0pt of rept_sqrt] (sqrt) {\phantom{xxx}sqrtsd xmm0, xmm0} ;
\node[codeblock,fill=delay_bg,below=0pt of sqrt] (cvtsd2si) {cvtsd2si rax, xmm0} ;
\node[codeblock,fill=read_bg,below=0pt of cvtsd2si] (read1) {mov r13, [r15]} ;
\node[codeblock,fill=fill_free,font=\sffamily\itshape,below=0pt of read1] (fill_rept) {repeat $k$ times:} ;
\node[codeblock,fill=fill_free,below=0pt of fill_rept] (fill) {\phantom{xxx}imul rax, 3} ;
\node[codeblock,fill=read_bg,below=0pt of fill] (read2) {mov r14, [r15]} ;

\draw[reorder_arrow] (read2.west) to[out=180,in=180,looseness=0.75] node[reorder_caption] {reorder\\\texttt{r14}$=$\texttt{r13}} (read1.south west) ;

\node[block_label] at (mfence.east) {1} ;
\node[block_label] at (rept_sqrt.east) {2} ;
\node[block_label] at (read1.east) {3} ;
\node[block_label] at (fill.north east) {4} ;
\node[block_label] at (read2.east) {5} ;
\end{tikzpicture}
    }
    \caption{Scheduler capacity $k$ not exceeded: The second bingo read~\ding{186} is reordered and retrieves the same bingo number as the first read~\ding{184}.\label{suf:bingo-no-contention}}
  \end{subfigure}\hspace{0.05\hsize}
  \begin{subfigure}[t]{0.45\hsize}
    \resizebox{\hsize}{!}{
      \begin{tikzpicture}
\definecolor{blue1}{HTML}{00aaff}
\definecolor{blue2}{HTML}{007fd0}
\colorlet{drain_bg}{blue2!10}
\colorlet{delay_bg}{blue1!30}
\colorlet{delay_rept_bg}{blue1!50}
\colorlet{read_bg}{orange!30}
\colorlet{fill_free}{green!20}
\colorlet{fill_full}{red!40}
\tikzset{
    codeblock/.style={
        anchor=north,
        align=left,
        font=\ttfamily,
        minimum width=4.5cm,
        minimum height=0.6cm,
        text width=4cm,
        line width=0pt,
        inner ysep=0pt
    },
    stall_line/.style={
        ultra thick,
        red
    },
    stall_caption/.style={
        align=center,
        inner sep=0px,
        right
    },
    block_label/.style={
        circle,
        draw=none,
        fill=black,
        text=white,
        font=\bfseries\footnotesize,
        inner sep=0.3mm,
        minimum width=6.3mm
    }
}

\path (-4.5,0.25) rectangle (3,-6.75) ;

\node[codeblock,fill=drain_bg] at (0,0) (mfence) {mfence} ;
\node[codeblock,fill=delay_bg,below=0pt of mfence] (mov_rax) {mov rax, 12345678} ;
\node[codeblock,fill=delay_bg,below=0pt of mov_rax] (cvtsi2sd) {cvtsi2sd xmm0, rax} ;
\node[codeblock,fill=delay_rept_bg,font=\sffamily\itshape,below=0pt of cvtsi2sd] (rept_sqrt) {repeat 12 times:} ;
\node[codeblock,fill=delay_rept_bg,below=0pt of rept_sqrt] (sqrt) {\phantom{xxx}sqrtsd xmm0, xmm0} ;
\node[codeblock,fill=delay_bg,below=0pt of sqrt] (cvtsd2si) {cvtsd2si rax, xmm0} ;
\node[codeblock,fill=read_bg,below=0pt of cvtsd2si] (read1) {mov r13, [r15]} ;
\node[codeblock,fill=fill_free,font=\sffamily\itshape,below=0pt of read1] (fill_rept) {repeat $k$ times:} ;
\node[codeblock,fill=fill_free,below=0pt of fill_rept] (fill_free) {\phantom{xxx}imul rax, 3} ;
\node[codeblock,fill=fill_full,below=0pt of fill_free] (fill_full) {imul rax, 3} ;
\node[codeblock,fill=read_bg,below=0pt of fill_full] (read2) {mov r14, [r15]} ;

\draw[stall_line] (fill_full.north west) +(-2.2,0) node[stall_caption] {Pipeline stall\\\texttt{r14}$\neq$\texttt{r13}} -- (fill_full.north east) -- +(0.25,0) ;

\node[block_label] at (mfence.east) {1} ;
\node[block_label] at (rept_sqrt.east) {2} ;
\node[block_label] at (read1.east) {3} ;
\node[block_label] at (fill_free.north east) {4a} ;
\node[block_label] at (fill_full.east) {4b} ;
\node[block_label] at (read2.east) {5} ;
\end{tikzpicture}
    }
    \caption{Scheduler capacity $k$ exceeded: Contention causes a pipeline stall, hence the second read~\ding{186} is not reordered and retrieves a different bingo number than the first read~\ding{184}.\label{suf:bingo-contention}}
  \end{subfigure}
  \caption{Measuring scheduler contention with a bingo race. After draining the pipeline~\ding{182}, we fill the scheduler queue with repetitions of the priming instruction~\ding{185}, delayed by a high-latency input operand dependency chain~\ding{183}. The bingo variable at \texttt{[r15]} is constantly updated by the bingo thread on another core. If the pipeline stalls due to scheduler contention, \texttt{r14} will contain a different value than \texttt{r13}.}
  \label{fig:squip-bingo}
\end{figure}

\section{The Accuracy of the Measurement}\label{sec:measurement}
With this section we address \textbf{RQ2}.
Our initial experiments show that the two measurements methods, performance counters and non-serialized hardware timer reads, described by Gast~\etal\cite{Gast2023SQUIP} yield imprecise results.
Their measurements with both methods show a scheduler capacity of 22 instead of the documented 24 \uops.
Our hypothesis is that the complex \texttt{rdpru} and \texttt{rdtsc} instructions use multiple scheduler entries for themselves, influencing the measurement.
Additionally, our JavaScript covert channel in \cref{sec:covert} requires a measurement method that only uses instructions emitted by the JavaScript JIT compiler.
We therefore develop a novel method, using a race condition between a read from a bingo variable and the targeted scheduler queue.
With this method we measure the exact scheduler 1 queue size of 24 entries on Zen 3 and Zen 4.

\subsection{Measurements using an Out-of-Order Bingo Race Condition}\label{sub:bingothread}
Gast~\etal\cite{Gast2023SQUIP} use timer reads to detect if the CPU executes instructions out of order.
We show that this is also possible with a bingo variable that is modified by a thread running in parallel and that this yields a more precise result.\footnote{Our approach is named after the game Bingo, where numbers are openly announced, and matching numbers have a special meaning. Likewise, in our case, matching bingo numbers have a special meaning, namely that the scheduler queue was not full.}

In~\Cref{fig:squip-bingo}, we replace the hardware timer reads (\ie \texttt{rdpru}) with loads from a bingo variable at a specific memory address (in \texttt{r15}).
The bingo thread constantly updates this variable with new bingo numbers.
We have no requirements on the bingo numbers except for them to be frequently updated, without frequent value repetitions, \eg a (pseudo-random) number sequence.
In contrast to the timer reads~\cite{Gast2023SQUIP}, the \texttt{mov} instruction to read the bingo variable is decoded into only a single \uop~\cite{AMD2020OptimizationEPYC7003}, minimizing its influence on the measurement while still subjected to out-of-order execution.
The frequent updates keep the bingo variable in the L1 cache, preventing delays from memory accesses.

The second bingo access is only executed out-of-order when there is no contention on the targeted scheduler queue.
Thus, contention is detected based on whether the second read of the bingo variable is executed out-of-order or not.
If it is executed out-of-order, the second read of the bingo variable will retrieve the exact same bingo number, otherwise, it will retrieve a new, different bingo number.
In other words, when the queue is \emph{full} and the bingo variable read is about to be enqueued, the pipeline stalls and subsequent \uops are not reordered, \ie they do not reach their respective scheduler queues and thus the second bingo variable read does not succeed out of order.
With all measurement methods, there can be spurious stalls, which we investigate further in~\cref{sub:reduced-capacity}.

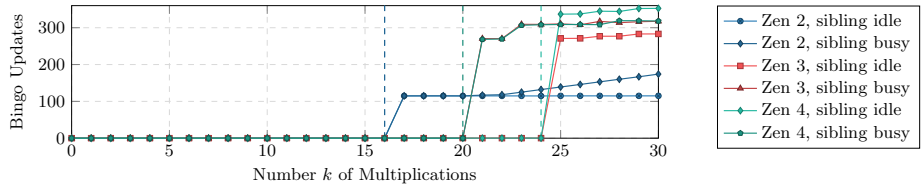
\begin{figure}[t]
  \resizebox{\hsize}{!}{
  \begin{tikzpicture}
    \begin{axis}[
        width=\hsize, 
        height=4cm,
        grid=major,
        grid style={dashed,gray!30},
        ticklabel style={
          /pgf/number format/.cd, fixed
        },
        xlabel=Number $k$ of Multiplications,
        ylabel=Bingo Updates,
        xmin=0,
        xmax=30,
        ymin=0,
        ymax=360,
        legend cell align={left},
        legend style={at={(1.1, 1)}, anchor=north west},
        xtick distance=5,
        scaled y ticks=false
      ]
      \addplot+[blue,semithick,mark=otimes*,mark size=0.5mm,mark options={draw=blue!60!black,fill=blue,solid,thin}]
        table[x=Number of multiplications,y=Zen 2,col sep=comma]{bingo_imul.csv};
      \addplot+[blue!80!black,semithick,mark=diamond*,mark size=0.6mm,mark options={draw=blue!50!black,fill=blue!80!black,solid,thin}]
        table[x=Number of multiplications,y=Zen 2 loop,col sep=comma]{bingo_imul.csv};
      \addplot+[wine,semithick,mark=square*,mark size=0.5mm,mark options={draw=wine!70!black,fill=wine,solid,thin}]
        table[x=Number of multiplications,y=Zen 3,col sep=comma]{bingo_imul.csv};
      \addplot+[wine!80!black,semithick,mark=triangle*,mark size=0.6mm,mark options={draw=wine!50!black,fill=wine!80!black,solid,thin}]
        table[x=Number of multiplications,y=Zen 3 loop,col sep=comma]{bingo_imul.csv};
      \addplot+[ocean,semithick,mark=diamond*,mark size=0.6mm,mark options={draw=ocean!70!black,fill=ocean,solid,thin}]
        table[x=Number of multiplications,y=Zen 4,col sep=comma]{bingo_imul.csv};
      \addplot+[ocean!80!black,semithick,solid,mark=pentagon*,mark size=0.5mm,mark options={draw=ocean!50!black,fill=ocean!80!black,solid,thin}]
        table[x=Number of multiplications,y=Zen 4 loop,col sep=comma]{bingo_imul.csv};
      \legend{Zen 2, sibling idle\\Zen 2, sibling busy\\Zen 3, sibling idle\\Zen 3, sibling busy\\Zen 4, sibling idle\\Zen 4, sibling busy\\}
      \addplot[samples=50, smooth,domain=0:16,blue,dashed,semithick] coordinates {(16,0)(16,360)};
      \addplot[samples=50, smooth,domain=0:16,blue!80!black,dashed,semithick] coordinates {(16,0)(16,360)};
      \addplot[samples=50, smooth,domain=0:24,ocean,dashed,semithick] coordinates {(24,0)(24,360)};
      \addplot[samples=50, smooth,domain=0:20,ocean!80!black,dashed,semithick] coordinates {(20,0)(20,360)};
    \end{axis}
  \end{tikzpicture}
  }
  \caption{Average load delay for different lengths $k$ of the multiplication block for Zen 2, 3 and 4; with sibling thread being busy or idle. ($n=\SIx{100000}$).}
  \label{fig:bingo-imul}
\end{figure}

\subsection{Evaluation}
We evaluate our bingo race measurement on an AMD Ryzen 7 3700X CPU (Zen 2), an AMD Zen 3 Ryzen 7 5800X CPU (Zen 3), and an AMD Ryzen 7 7700X CPU (Zen 4) with a varying number of multiplications $k$.
For each $k$, we repeat the measurement \SIx{100000} times, and track the average number of updates of the bingo number.
We avoid interference between the bingo thread and the measurement thread by pinning each of them to separate cores.

\Cref{fig:bingo-imul} shows the increased precision of our method.
On Zen~2, we measure a queue capacity of 16~\uops for scheduler~1, exactly matching the documentation~\cite{AMD2020OptimizationEPYC7002} and previous work~\cite{Gast2023SQUIP}.
On Zen~3 and Zen~4, we measure a capacity of 24~\uops, exactly matching the capacity published by AMD~\cite{AMD2020WhereGamingBegins} and two \uops more than reported by Gast~\etal\cite{Gast2023SQUIP}.
We further analyze this difference in~\cref{sub:reduced-capacity}.
Running an empty loop on the other hardware thread of the same core yields a reduced capacity of 20~\uops on Zen~3 and Zen~4, whereas there is no reduction on Zen~2.
This reveals the watermark mechanism on Zen~3 and Zen~4, in line with previous work~\cite{Gast2023SQUIP}.
Our results for more instructions are shown in \cref{tab:queue-instructions}.

\section{Detecting User Behavior via Scheduler Contention}\label{sec:keystrokes}

In this section, we demonstrate detection of singular non-repeatable events via scheduler contention, addressing \textbf{RQ3}.
We describe an attack to recover inter-keystroke timings from the X server and evaluate it on all integer execution unit schedulers, also extending our insights towards \textbf{RQ1}.
Our evaluation shows successful keystroke spying, with an F$_1$ score \SI{\geq 99.5}{\percent}.
We observe that the average deviation from the ground truth for correctly detected inter keystroke-timings is below \SI{4}{\milli\second}, indicating the high level of precision in our attack.

\subsection{Threat Model and Experimental Setup}
The attacker wants to recover inter-keystroke timings from a user entering their password into Gnome Display Manager (gdm3, the default login manager on Ubuntu) on a multi-user machine.
We assume that the attacker can execute native code on the target machine and has achieved co-location to the X server, \ie attack code and X server run on the same core but on different hardware threads.
We evaluate our attack on an AMD Ryzen 7 5800X CPU (Zen 3) running Ubuntu 20.04.6 LTS with X server 1.20.13 and gdm3 3.36.3.

\subsection{Inter-Keystroke Timing Attack}
Our attack exploits the watermark mechanism of the scheduler queues to detect bursts of high activity of the co-located X server.
The attack consists of an observation phase and a postprocessing phase.
First, the observer process repeatedly samples any of the scheduler queues to infer if the X server is active, recording timestamps each time it detects activity.
In the postprocessing phase, we group the timestamps to obtain the start time of each activity burst.

\paragrabf{Observation phase}
The observer process is co-located to the X server, which provides graphical services and handles user input.
User input, such as pressing a key, results in an activity burst.
Conversely, without user input, the X server remains predominantly idle, causing only low activity.
Since the scheduler queues are shared, an activity burst of the X server can be detected by the observer.

The observer continuously checks whether it can exceed the watermark limit on the targeted scheduler queue without causing a back-end stall.
If the available capacity is \emph{above} the watermark limit, the X server is currently \emph{idle}.
If the available capacity is \emph{below} the watermark limit, this indicates that the X server is \emph{busy} performing tasks like updating the display, handling mouse movements, or, notably, processing \emph{keystrokes}.
In these instances, the observer records a timestamp, resulting in a chronological list when the X server was active.

\begin{figure}[t]
  \resizebox{\hsize}{!}{
  \begin{tikzpicture}
    \begin{axis}[
        width=0.9\hsize,
        height=3cm,
        grid=major,
        grid style={dashed,gray!30},
        ticklabel style={
          /pgf/number format/.cd, fixed
        },
        xlabel=Time (\si{\milli\second}),
        ymin=0,
        ymax=2,
        ytick={0.2,1,1.8},
        yticklabels={key pressed,activity observed,keystroke detected},
      ]
      \addplot+[blue,only marks,mark=triangle,mark size=1.7mm,mark options={very thick}]
        table[x index=0,y expr=0.2,col sep=comma] {postprocess-example.csv} ;
      \addplot+[orange!50!black,only marks,mark=|,mark size=1.7mm,mark options={thick}]
        table[x index=1,y expr=1,col sep=comma] {postprocess-example.csv} ;
      \addplot+[red,only marks,mark=+,mark size=1.7mm,mark options={very thick}]
        table[x index=2,y expr=1.8,col sep=comma] {postprocess-example.csv} ;
    \end{axis}
  \end{tikzpicture}
  }
  \caption{
Keystroke (\textcolor{blue}{$\mathbf{\triangle}$}) timings observed via scheduler 3 by a co-located observer due to the watermark mechanism (\textcolor{orange!50!black}{$\mid\mid\mid$}). Multiple samples with reduced scheduler capacity are clustered and filtered, resulting in a clear recovered keystroke signal (\textcolor{red}{+}).}
  \label{fig:ks-postprocess}
\end{figure}
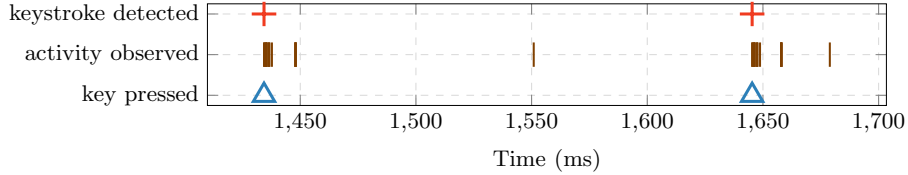

\paragrabf{Postprocessing}
A single keystroke triggers X server activity over several sampling intervals, causing the observer to record multiple timestamps, as illustrated in \cref{fig:ks-postprocess}.
The postprocessing phase filters these timestamps in two steps:
In the first step, we group them into clusters.
We consider two subsequent timestamps to be part of the same cluster if their difference is below or equal a threshold $\tidle$.
In the second step, we filter out clusters with less than $\smin$ samples, removing noise caused by X server activity unrelated to user input.
The first timestamp in each remaining cluster is the assumed point in time of a keystroke.

\cref{fig:ks-postprocess} exemplifies the postprocessing phase, where the observed activity clusters (\textcolor{orange!50!black}{$\mid\mid\mid$}) in the timer intervals 1434-1448 and 1645-1659 are interpreted as keystrokes (\textcolor{red}{+}) at timestamps 1434 and 1645, respectively.
Conversely, the observed activities at timestamps 1551 and 1679 are filtered out as noise.

\begin{table}[t]
  \caption{Inter-keystroke Timing Attack Evaluation}
  \scriptsize
  \label{tab:ks-eval}
  \resizebox{\hsize}{!}{
    \begin{tabular}{lrrrr}
      \toprule
                                                & Scheduler 0   & Scheduler 1   & Scheduler 2   & Scheduler 3   \\
      \midrule
        Recorded Keystrokes                     &           100 &           100 &           100 &           100 \\
        False negatives                         &             0 &             0 &             1 &             0 \\
        False positives                         &             1 &             0 &             0 &             0 \\
        $F_1$ score                             &         $0.995$ &         $1.000$ &         $0.995$ &         $1.000$ \\
        Standard deviation (\unit{\milli\second}) & $2.555$ ($n=99$)  & $3.244$ ($n=100$) & $2.307$ ($n=99$)  & $0.901$ ($n=100$) \\
      \bottomrule
    \end{tabular}
  }
\end{table}

\subsection{Evaluation}
We evaluate our attack on all four integer execution unit schedulers as follows:

Using a programmable USB keyboard emulator, we inject 100 keystrokes into the gdm3 password prompt.
Between each pair of keystrokes, we insert a delay between \SI{150}{\milli\second} and \SI{300}{\milli\second}, simulating the typing speed of a skilled typist~\cite{Pinet2016}.
The keyboard emulator records the actual keystroke timings as the ground truth.
We post-process ($\tidle = \SI{0.1}{ms}$, $\smin = 10$) the recorded timestamps, resulting in a trace of keystroke times.
The values for $\tidle$ and $\smin$ were found empirically to achieve good filtering for traces recorded on all integer scheduler queues.

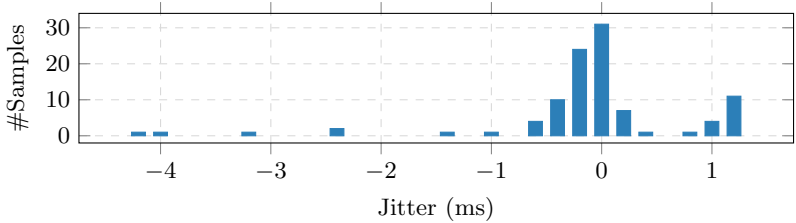
\begin{figure}[t]
  \centering
  \begin{tikzpicture}
    \begin{axis}[
        ybar,
        bar width=0.125,
        width=0.9\hsize,
        height=3.3cm,
        grid=major,
        grid style={dashed,gray!30},
        ticklabel style={
          /pgf/number format/.cd, fixed
        },
        xlabel=Jitter (\si{\milli\second}),
        ylabel=\#Samples
      ]
      \addplot+[blue]
        table[x index=0,y index=1,col sep=comma] {keystrokes-sched3.csv} ;
    \end{axis}
  \end{tikzpicture}
  \caption{
The jitter of the individual recorded keystrokes (from the aligned ground truth), observed via scheduler 3.
The low number of outliers and high concentration around \SI{0}{\milli\second} shows that inter-keystroke timings are extracted with very high accuracy.
}
  \label{fig:keystroke-jitter}
\end{figure}

For inter-keystroke timing attacks, the exact timing between individual key\-strokes is the crucial information.
Due to other system activity, the measured inter-keystroke timings can vary slightly, making the attack, and possibly recovered keys, less precise.
These slight inaccuracies are called jitter and should be as low as possible.
We compute the jitter of our measurement as follows:

We first align the trace with the ground truth by computing the mean squared error between the ground truth and the signal shifted by $\Delta t$ and minimizing it.
We then check for false negatives and false positives against the ground truth and count them, giving us the numbers in \cref{tab:ks-eval}.
The very low numbers of 1 or less are possible due to the filtering with $\tidle$ and $\smin$.
To compute the jitter, we remove false negatives and positives from the signal.
For each of the remaining true positive keystroke timings, we compute the distance to the ground truth.
\cref{fig:keystroke-jitter} shows the distribution of the jitter of 100 recorded keystrokes.
The small number of outliers and high concentration around \SI{0}{\milli\second} indicate a very low jitter.
The high accuracy of our attack is summarized in \cref{tab:ks-eval}.
For all scheduler queues, we obtain $F_1$ scores close to 1, showing reliable keystroke detection.
The standard deviation is below \SI{1}{\milli\second} for scheduler queue 3 and maximum \SI{3.244}{\milli\second} for scheduler queue 1, indicating very precise timings.

As we attack the login screen, before the user session is started, the only sources of X server activity are password keystrokes and mouse movements.
Mouse movements become long bursts in the signal, easily filtered out.
While typing, users typically have both hands on the keyboard without using the mouse.

\section{JavaScript-based Scheduler Contention Covert Channel}\label{sec:covert}
In this section, we show that scheduler contention can exploited from a website, addressing \textbf{RQ4}.
We design and evaluate a covert channel using JavaScript in Firefox 114, similar to other state-of-the-art works~\cite{Rokicki2022webport,VanSchaik2019RIDL,Schwarz2017Timers,Vila2017,Oren2015}.
We achieve a raw transmission rate of \SI{1}{\kilo\bit\per\second} with an error rate of less than \SI{1.8}{\percent} in a cross-browser-window setting, bypassing cross-origin policies and site isolation.

\subsection{Threat Model and Experimental Setup}
We use a Ryzen 7 5800X (Zen 3) and a 7700X (Zen 4), 8 cores and 16 SMT threads each, with a default Ubuntu 20.04 and Firefox 114.
Like other browser-based covert channels~\cite{Rokicki2022webport,VanSchaik2019RIDL,Schwarz2017Timers,Vila2017,Oren2015}, we assume receiver and sender have no other communication channel, \eg due to cross-origin policies and site isolation.
We make no assumptions on CPU frequency or co-location of sender and receiver.

\subsection{Scheduler Contention in JavaScript}
Compared to microarchitectural attacks from native code, browser-based attacks have to overcome additional challenges.
In particular, to induce and observe scheduler contention across multiple browser instances, these are:
\begin{itemize}
\item[\textbf{C1}] Find a scheduler that can be targeted from JavaScript, with instructions that the JIT compiler produces.
\item[\textbf{C2}] Craft code that generates a tight and non-bloated sequence of priming instructions for the targeted scheduler, \ie surrounding code must not interfere with other schedulers or prevent the contention of the targeted scheduler.
\item[\textbf{C3}] Measure contention without access to hardware timers.
\item[\textbf{C4}] Achieve co-location between sender and receiver.
\end{itemize}
We solve each challenge and then combine them to our end-to-end covert channel.

\paragrabf{C1} Up to this point, priming a scheduler required fine-tuned assembly code with carefully chosen instructions.
JavaScript not providing direct control over the machine code makes priming more difficult, hence the targeted scheduler must be carefully chosen:
Scheduler 0 is heavily used by several instructions, increasing the noise and reducing the transmission rate.
Scheduler 1 is best primed with \texttt{imul}, which is easy to emit from JavaScript via \texttt{Math.imul}.
While we found instructions to prime scheduler 2 and 3 in~\cref{sec:queues}, we have not found these specialized string operations in code emitted by the JIT compiler.
The FPU schedulers can only be targeted at once, with a high number of operations, and are also heavily used, as all JavaScript numbers are floating point numbers by default~\cite{Mozilla2023Number}.
This leaves scheduler 1 as the most plausible choice, solving \textbf{C1}.

\paragrabf{C2} We use asm.js (\eg \texttt{|0} or the unary \texttt{+}) to tell the JIT compiler to use integer data types~\cite{Mozilla2023AsmJS,Herman2014AsmJS}.
These type hints enable the compiler to omit type checks and to directly emit machine instructions suitable for the hinted data type.
We prime scheduler 1 with a chain of \SIx{20} dependent calls to \texttt{Math.imul}, delayed by a chain of \SIx{12} calls to \texttt{Math.sqrt} (see \cref{app:javasquip}, \cref{lst:javasquip}).
To prevent the compiler from over-optimizing, we pseudo-randomize the initial value for the square roots and one operand of the multiplications.
Disassembling the JIT-compiled code shows that our instruction sequence is compiled into a chain of \SIx{12} \texttt{sqrtsd}, followed by \SIx{20} \texttt{imul} (see \cref{lst:javasquip-disasm} in \cref{app:javasquip}).

\paragrabf{C3} Gast~\etal\cite{Gast2023SQUIP} measured scheduler contention via non-serialized hardware timer reads, which are not available in JavaScript.
However, our bingo race approach (\cref{sub:bingothread}) eliminates that requirement, solving \textbf{C3}.
Like other state-of-the-art microarchitectural attacks from JavaScript~\cite{Rokicki2022webport,Schwarz2017Timers,Lipp2017practical,Gras2017aslr}, this requires a Web Worker with a \texttt{SharedArrayBuffer}.
The Web Worker constantly updates the bingo variable within the shared memory.
The receiver reads the bingo variable using \texttt{Atomics.load}.
Each read from the bingo variable is JIT-compiled into a single \texttt{mov r8d, [rsi+rbp*4]} instruction.
This instruction immediately follows the last multiplication, allowing it to be executed out-of-order when the scheduler is not full.

\begin{figure}[t]
  \centering
  \begin{subfigure}[t]{0.3\hsize}
    \resizebox{\hsize}{!}{
      \begin{tikzpicture}
    \colorlet{high_fg}{white}
    \colorlet{high_bg}{red!80}
    \colorlet{low_fg}{black}
    \colorlet{low_bg}{green!20}

    \tikzset{
        pics/core/.style args={#1/#2/#3/#4/#5/#6}{
            code={
                \draw[very thick] (0,-0.2) rectangle (1,1.2) ;

                \draw[fill=#3] (0,0.5) rectangle (1,1.2) ;
                \node[text=#2,font=\bfseries\LARGE] at (0.5,0.85) {#1} ;

                \draw[fill=#6] (0,-0.2) rectangle (1,0.5) ;
                \node[text=#5,font=\bfseries\LARGE] at (0.5,0.15) {#4} ;
            }
        }
    }

    \draw (0.0,0) pic{core=R/high_fg/high_bg/R/high_fg/high_bg} ;
    \draw (1.75,0) pic{core=R/high_fg/high_bg/R/high_fg/high_bg} ;
    \draw (3.5,0) pic{core=S 0/low_fg/low_bg/R/high_fg/high_bg} ;
    \draw (5.25,0) pic{core=R/high_fg/high_bg/B/low_fg/low_bg} ;

    \node[circle,fill=black,text=white,font=\bfseries\Large,inner sep=0.8mm] at(3.45,0.15) {1} ;
    \node[circle,fill=black,text=white,font=\bfseries\Large,inner sep=0.8mm] at(5.20,0.85) {2} ;
\end{tikzpicture}
    }
    \caption{Sending a `0'}\label{suf:colocation-0}
  \end{subfigure}\hspace{0.03\hsize}
  \begin{subfigure}[t]{0.3\hsize}
    \resizebox{\hsize}{!}{
      \begin{tikzpicture}
    \colorlet{high_fg}{white}
    \colorlet{high_bg}{red!80}
    \colorlet{low_fg}{black}
    \colorlet{low_bg}{green!20}

    \tikzset{
        pics/core/.style args={#1/#2/#3/#4/#5/#6}{
            code={
                \draw[very thick] (0,-0.2) rectangle (1,1.2) ;

                \draw[fill=#3] (0,0.5) rectangle (1,1.2) ;
                \node[text=#2,font=\bfseries\LARGE] at (0.5,0.85) {#1} ;

                \draw[fill=#6] (0,-0.2) rectangle (1,0.5) ;
                \node[text=#5,font=\bfseries\LARGE] at (0.5,0.15) {#4} ;
            }
        }
    }

    \draw (0.0,0) pic{core=R/high_fg/high_bg/R/high_fg/high_bg} ;
    \draw (1.75,0) pic{core=R/high_fg/high_bg/B/low_fg/low_bg} ;
    \draw (3.5,0) pic{core=R/high_fg/high_bg/R/high_fg/high_bg} ;
    \draw (5.25,0) pic{core=S 1/high_fg/high_bg/R/high_fg/high_bg} ;

    \node[circle,fill=black,text=white,font=\bfseries\Large,inner sep=1mm] at(1.65,0.85) {1} ;
\end{tikzpicture}
    }
    \caption{Sending a `1'}\label{suf:colocation-1}
  \end{subfigure}\hspace{0.03\hsize}
  \begin{subfigure}[t]{0.3\hsize}
    \resizebox{\hsize}{!}{
      \begin{tikzpicture}
    \colorlet{high_fg}{white}
    \colorlet{high_bg}{red!80}
    \colorlet{low_fg}{black}
    \colorlet{low_bg}{green!20}

    \tikzset{
        pics/core/.style args={#1/#2/#3/#4/#5/#6}{
            code={
                \draw[very thick] (0,-0.2) rectangle (1,1.2) ;

                \draw[fill=#3] (0,0.5) rectangle (1,1.2) ;
                \node[text=#2,font=\bfseries\LARGE] at (0.5,0.85) {#1} ;

                \draw[fill=#6] (0,-0.2) rectangle (1,0.5) ;
                \node[text=#5,font=\bfseries\LARGE] at (0.5,0.15) {#4} ;
            }
        }
    }

    \draw (0.0,0) pic{core=R/high_fg/high_bg/R/high_fg/high_bg} ;
    \draw (1.75,0) pic{core=S 0/low_fg/low_bg/B/low_fg/low_bg} ;
    \draw (3.5,0) pic{core=R/high_fg/high_bg/R/high_fg/high_bg} ;
    \draw (5.25,0) pic{core=R/high_fg/high_bg/R/high_fg/high_bg} ;
\end{tikzpicture}
    }
    \caption{Sender co-located with bingo thread}\label{suf:colocation-fail}
  \end{subfigure}
  \begin{subfigure}{0.95\hsize}
    \resizebox{\hsize}{!}{
      \begin{tikzpicture}
    \colorlet{high_fg}{white}
    \colorlet{high_bg}{red!80}
    \colorlet{low_fg}{black}
    \colorlet{low_bg}{green!20}

    \draw[fill=high_bg] (0,2) rectangle +(0.5,-0.5) ;
    \node[anchor=west] at (0.5,1.72) {thread causing \textbf{high} contention level};
    \draw[fill=low_bg] (7,2) rectangle +(0.5,-0.5) ;
    \node[anchor=west] at (7.5,1.72) {thread causing \textbf{low} contention level};
\end{tikzpicture}
    }
  \end{subfigure}
  \caption{Spawning multiple receiver threads to achieve co-location. Each core has two hardware threads: one occupied by the bingo thread (B) the other by the sender (S). Other hardware threads run receiver threads (R), measuring scheduler contention. The transmitted bit is recovered from the number of threads that have observed a low contention level (a, b), unless the sender is co-located with the bingo thread (c).}\label{fig:colocation}
\end{figure}
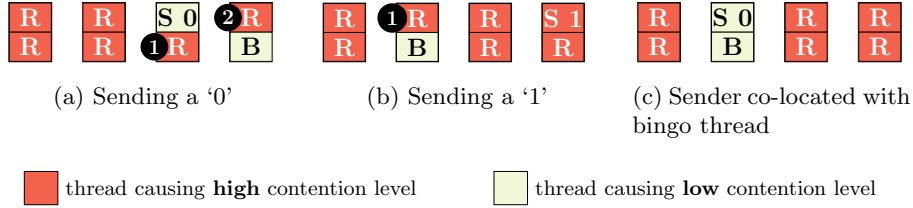

\paragrabf{C4} To observe scheduler contention across processes, threads must be co-located on a physical core.
As JavaScript cannot control this, prior work~\cite{Rokicki2022webport} ran multiple sender instances.
If the number of tasks matches the number of hardware threads, this results in a one-to-one assignment of tasks to individual hardware threads, with inert core affinities.
We instead run multiple receiver Web Workers, leaving one for the sender and one for the bingo thread.

Consequently, the majority of the receiver workers is co-located to another receiver, as shown in~\cref{fig:colocation}.
These receivers observe a high level of scheduler contention, as both receivers continuously prime and probe the shared scheduler queue.
Typically, one of the receivers is co-located with the bingo thread and one of them is co-located with the sender.
A receiver co-located with the bingo thread observes a low contention level, as the bingo thread does not perform integer multiplications.
A receiver co-located with the sender observes either a low or a high contention level, depending on the bit that is transmitted.

However, it is possible that the OS co-locates the sender with the bingo thread, see \cref{suf:colocation-fail}.
In this case, each receiver is co-located to another receiver and, therefore, observes high contention levels.
This case is trivial to detect and the attacker can just restart the communication until the threads are distributed in a different way.
Hence, this approach solves the last challenge, \textbf{C4}.

\subsection{Time-sliced Bit Transmission}
Our covert channel continuously transmits the same bit over a time slice, with a shared clock for synchronization.
We use the coarse-grained \texttt{Date.now} for start and end of each time slice but not for any measurements, as \texttt{Date.now} has side effects (possibly triggering a serializing system call) that prohibit use during out-of-order execution.
The minimum time slice is limited by the resolution of \texttt{Date.now}, which is \SI{1}{\milli\second} on Firefox 114, \ie the raw transmission rate is \SI{1}{\kilo\bit\per\second}.
To transmit a `0', the sender repeatedly checks \texttt{Date.now} to wait for the end of the time slice.
To transmit a `1', the sender additionally executes a chain of dependent \texttt{Math.imul} calls to cause scheduler contention.

Each receiver continuously measures contention on scheduler 1.
Every millisecond, one of the receivers collects and evaluates the results from all receivers.
As stated previously, most receivers measure a constantly high contention level, one measures no contention and one the actual signal from the sender, which is the second-lowest average contention level.
If this receiver's contention level exceeds a predefined threshold, we consider the retrieved bit a `1' (see~\cref{suf:colocation-1}); otherwise, we consider the retrieved bit a `0' (see~\cref{suf:colocation-0}).

\subsection{Evaluation}
Our covert channel is around 5 times faster than other unmitigated state-of-the-art covert channels in the browser (\cf~\cref{tab:covert-compare}).
In contrast to many of these works, our covert channel does not rely on high-overhead operations, such as intra-browser message passing~\cite{Vila2017}, cache evictions~\cite{Oren2015,Schwarz2017Timers,VanSchaik2019RIDL} or disk accesses~\cite{VanGoethem2017isolated}.
Additionally, the back-end stalls caused by scheduler contention enable our receiver to easily distinguish contention and non-contention cases, even from within a sandboxed JavaScript environment.

We evaluate our covert channel by transmitting 15 random messages, each \SI{5000}{\byte} in size, each divided into 10 packets.
For AMD Zen 3, we detected 17 lost packets, out of 150, due to the sender being co-located with the bingo thread.
We achieved a bit error rate of \CovertErrorZenThree.
With a raw bandwidth of \SI{1}{\kilo\bit\per\second}, this results in a true capacity of \CovertTrueCapacityZenThree.

For AMD Zen 4, we transmit 10 random messages, \SI{5000}{\byte} in size, divided into 10 packets.
Out of 100 packets, 4 packets were lost due to the sender being co-located with the bingo thread.
The average bit error rate is \CovertErrorZenFour, with a true capacity of \CovertTrueCapacityZenFour.
Based on our findings, we conclude that JavaScript-based scheduler contention side-channel attacks are practical.
Additionally, we highlight that the attack surface created by these attacks is larger than previously anticipated.

\section{Discussion}\label{sec:discussion}
Our results on Zen 4, non-repeatable events, and from JavaScript, show that scheduler contention side channels are more relevant than previously known.
Prior work focused on native attacks and non-constant-time RSA code~\cite{Gast2023SQUIP}.
Consequently, the mitigations they discuss focus primarily on this scenario and do not mitigate the attacks we presented.
First, generic constant-time code for, \eg user input, would incur extreme runtime and latency overheads, rendering systems completely unusable.
Second, using a watermark to implement strong non-interference is not trivial, especially with an attacker controlling the instruction stream (via JavaScript).
Third, disabling SMT has prohibitive performance overheads~\cite{Cutress2020SMT}.
Fourth, co-scheduling and core-scheduling have significant performance overheads~\cite{Faggioli2019CoreScheduling} in a range of \SIrange{8}{91}{\percent}, rendering these software-level mitigations unpractical against our attacks.
Using a single scheduler queue, like Intel CPUs, or a symmetric scheduler and execution unit design is also insufficient:
The FPU execution units have a symmetric scheduler design and we still were able to fill the entire scheduler queue including the non-scheduling queue; Intel CPUs, in turn, are susceptible to port contention attacks~\cite{Aldaya2018,Bhattacharyya2019,Rokicki2022webport}.

\paragrabf{Related Work}\label{sec:related_work}
Oren~\etal\cite{Oren2015} demonstrated the first microarchitectural side-channel attack in the browser, in the form of a website-fingerprinting attack and a covert channel with a reported raw transmission rate of \SI{320}{\kilo\bit\per\second}, using \PrimeProbe.
As the high-resolution timer they used is no longer available on current browsers, several works demonstrated attacks without a high-resolution timer~\cite{Schwarz2017Timers,Shusterman2019,Shusterman2021Prime}.
Lipp~\etal\cite{Lipp2017practical} demonstrated interrupt-timing attacks in JavaScript, showcasing website and user fingerprinting, inter-keystroke timing attacks, and a covert channel.
Schwarz~\etal\cite{Schwarz2017Timers} presented a JavaScript-based covert channel using DRAM row access timing differences.
Rokicki~\etal\cite{Rokicki2022webport} constructed a covert channel in WebAssembly based on port contention.
They achieved a true capacity of \SI{184}{\bit\per\second}, making it currently the fastest browser-based covert channel on up-to-date browsers.
With a true capacity of \CovertTrueCapacityZenFourShort, our covert channel is more than 5 times faster.
Other works exploit speculative data loads~\cite{VanSchaik2019RIDL}, disk contention~\cite{VanGoethem2017isolated} or memory throttling~\cite{Rushanan2016malloryworker} from JavaScript, or focus on power, frequency, and temperature side channels~\cite{Dipta2022DF,Taneja2023Hot}.
In addition to these attacks, targeting specific hardware aspects, side-channel attacks can also target the browser itself~\cite{Vila2017,Weinberg2011}.

Several works demonstrated timing attacks on user input, specifically in\-ter-key\-strokes, potentially recovering or reducing the entropy of passwords.
The seminal work by Song~\etal\cite{Song2001} inferred keystroke timings from SSH connections.
Ristenpart~\etal\cite{Ristenpart2009} presented a cross-VM cache timing attack on keystrokes.
Gruss~\etal\cite{Gruss2015Template} presented a semi-automated \FlushReload attack on keystrokes.
Other works exploited DRAM row accesses~\cite{Pessl2016}, interrupts~\cite{Lipp2017practical}, browser event loops~\cite{Vila2017} or the CPU frequency~\cite{Dipta2022DF} to infer inter-keystroke timings.

Keyboard input typically triggers interrupts.
Thus, interrupts have been exploited in several works, using \texttt{rdtsc} differences~\cite{Schwarz2018KeyDrown}, \PrimeProbe~\cite{Schwarz2018KeyDrown}, and \texttt{procfs} files~\cite{Schwarz2018KeyDrown,Zhang2009keystroke,Jana2012memento}.
We extend the state-of-the-art by demonstrating key\-stroke timing attacks via another microarchitectural element that would remain exploitable even when the other side channels are closed.

\section{Conclusion}\label{sec:conclusion}
In this paper, we extended the state-of-the-art on scheduler queue side channels in four directions, covering (1) all scheduler queues on (2) both Zen 3 and the new Zen 4 microarchitecture, (3) singular non-repeatable keystroke events, and (4) a pure JavaScript-based attack running in Firefox.
We showed that all scheduler queues can be exploited and introduced a new measurement method, based on a timingless out-of-order race condition.
We demonstrated keystroke timing attacks on all scheduler queues, with $F_1$ scores close to 1 and a standard deviation from the ground truth below \SI{1}{\milli\second} for scheduler queue 3 and at most \SI{3.244}{\milli\second} for scheduler queue 1.
As native code execution is not a requirement for scheduler queue side channels, the research for efficient and effective mitigations becomes a more urgent issue.
Our end-to-end JavaScript attack underlines this with covert data transmission across browser windows, bypassing cross-origin policies and site isolation, and achieving true capacities of \CovertTrueCapacityZenFourShort{} on Zen 4.

\section*{Acknowledgments}
We thank our anonymous reviewers for their valuable feedback and Martin Schwarzl for help with initial experiments.
This research is supported in part by the European Research Council (ERC project FSSec 101076409), the Austrian Science Fund (FWF SFB project SPyCoDe 10.55776/F85), and the Austrian Research Promotion Agency (FFG project SEIZE 888087).
Additional funding was provided by generous gifts from Red Hat, Google, Intel and AWS.
Any opinions, findings, and conclusions or recommendations expressed in this paper are those of the authors and do not necessarily reflect the views of the funding parties.

\bibliographystyle{splncs04}
\bibliography{main}

\FloatBarrier

\appendix
\crefalias{section}{appendix}

\section{Validation of the Bingo Race Approach}\label{app:coun}\label{sub:bingo-limits}
To verify that we indeed observe scheduler contention with our out-of-order bingo variable race condition, we replace the \texttt{imul} priming instruction in block~\ding{185} of~\cref{fig:squip-bingo} with \texttt{add rax, 3}, which can be executed by all the 4 ALUs of the Zen 2 and the Zen 3 microarchitectures~\cite{AMD2020OptimizationEPYC7002,AMD2020OptimizationEPYC7003}.
On Zen~2, we measure a capacity of 64~\uops, matching the documented total capacity of all the 4 schedulers that are connected to an ALU.
On Zen~3, we measure a capacity of 90~\uops, which is 6~\uops less than the documented total capacity ($4 \times 24 = 96$~\uops~\cite{AMD2020WhereGamingBegins}) of all the 4 integer execution unit schedulers on that machine, which is again 2~\uops more than measured by previous work~\cite{Gast2023SQUIP}.
We discuss the reason for the lower observed capacity in~\cref{sub:reduced-capacity}.

One drawback of this method is that the memory access ordering rules of the ISA prevent reordering of the second bingo variable read in some cases.
For a simple example, if we ensure \texttt{rdi} contains a valid but uncached address and replace the priming instruction with \texttt{imul rax, [rdi]}, then the second bingo variable read will not retire until \texttt{[rdi]} has been read for the multiplication.
We verify this by inserting a \texttt{clflush [rdi]} instruction before block~\ding{182} in \cref{fig:squip-bingo} and replacing the multiplication in block~\ding{185} with \texttt{imul rax, [rdi]}.
With this, we only observe the same bingo number twice if $k=0$, whereas for $k=1$ we immediately see \SIx{198.395} (Zen 2) and \SIx{443.839} (Zen 3) bingo number updates on average ($n=100\,000$).
This shows that the uncached memory access indeed delays the subsequent second bingo variable read, entirely overshadowing the delay caused by scheduler contention.
In a control experiment using \texttt{rdpru} instead of a bingo thread, we observe no such effect.
Also, if we keep the bingo thread and the memory operand, but remove the \texttt{clflush [rdi]} instruction, we observe the same capacity limits as with the original experiment, because the memory operand is cached and accessing it does not cause such a large delay.

For a second example, we ensure \texttt{rdi + rax} contains a valid, cached memory address after block~\ding{184} in \cref{fig:squip-bingo}.
We now initialize \texttt{r10} with a fixed value and replace the priming instruction in block~\ding{185} with \texttt{imul r10, [rdi+rax]}, so that the address of its memory operand depends on the long-latency calculation of \texttt{rax}.
We again observe that the second bingo variable read is delayed for $k>0$, as (per memory access order rules) now \texttt{[rdi + rax]} has to be read before the bingo variable, which in turn can only be done after the result for \texttt{rax} has been computed.

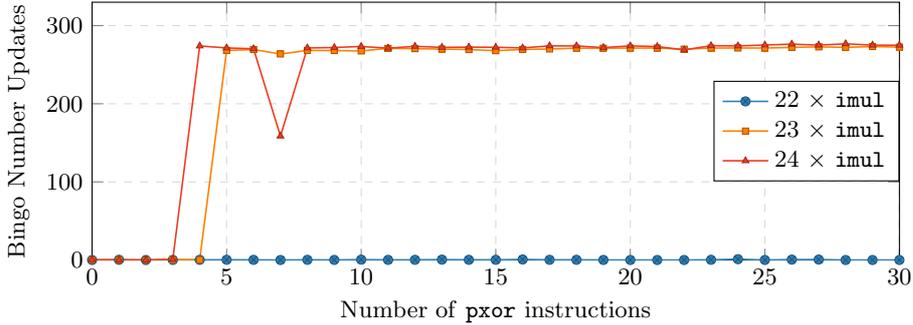
\begin{figure}[t]
  \begin{tikzpicture}
    \begin{axis}[
        width=\hsize,
        height=5cm,
        grid=major,
        grid style={dashed,gray!30},
        ticklabel style={
          /pgf/number format/.cd, fixed
        },
        xlabel=Number of \texttt{pxor} instructions,
        ylabel=Bingo Number Updates,
        xmin=0,
        xmax=30,
        ymin=0,
        ymax=330,
        legend cell align={left},
        legend style={at={(1, 0.5)}, anchor=east},
        scaled y ticks=false
      ]
      \addplot+[blue,semithick,mark=otimes*,mark size=0.6mm,mark options={draw=blue!70!black,thin,fill=blue,solid}]
        table[x=neutral,y=22 imul,col sep=comma]{spurious_stalls.csv};
      \addplot+[orange,semithick,mark=square*,mark size=0.4mm,mark options={draw=orange!70!black,thin,fill=orange,solid}]
        table[x=neutral,y=23 imul,col sep=comma]{spurious_stalls.csv};
      \addplot+[red,semithick,mark=triangle*,mark size=0.5mm,mark options={draw=red!70!black,thin,fill=red,solid}]
        table[x=neutral,y=24 imul,col sep=comma]{spurious_stalls.csv};
      \legend{22 \texttimes{} \texttt{imul}\\23 \texttimes{} \texttt{imul}\\24 \texttimes{} \texttt{imul}\\}
    \end{axis}
  \end{tikzpicture}
  \caption{Spurious back-end stalls on Zen 3. Even though the scheduler queue capacity is not exceeded, the CPU stalls a few instructions after the queue was primed with at least 23 multiplications.}
  \label{fig:spurious-stalls}
\end{figure}

\section{Spurious Scheduler Stalls}\label{sub:reduced-capacity}
Unlike earlier~\cite{Gast2023SQUIP} scheduler contention measurement methods, which used performance counters or non-serialized hardware timer reads via the \texttt{rdpru} instruction, our bingo method result in the exact capacity of scheduler 1 on Zen 3.
In this section, we show that the Zen 3 backend spuriously stalls after a few instructions, if the scheduler queue is primed with 23 or 24~\uops, explaining the 2~\uops difference between the actual capacity and the capacity observed with performance counters or \texttt{rdpru}.

First, we run the code in \cref{suf:bingo-no-contention} again with varying numbers of multiplications $k$, while monitoring the \texttt{IntSch1TokenStall} performance counter.
Like previous work~\cite{Gast2023SQUIP}, we observe a steep increase in stalled cycles with $k \ge 23$ multiplications, whereas the results from the bingo thread show that, with $k \le 24$ the backend does not stall before reaching the second bingo variable read.
This indicates that the backend might stall \emph{after} the second bingo variable read, even if the scheduler queue capacity is not exceeded.

We further investigate this with variations of \cref{suf:bingo-no-contention}, in which we prime the scheduler queue with fixed numbers of multiplications and insert a varying number of \texttt{pxor xmm1, xmm1} instructions between the multiplications and the second bingo variable read.
As the \texttt{pxor} instruction is handled by the FPU~\cite{AMD2020OptimizationEPYC7003}, it does not use the integer execution unit schedulers, however it adds a short delay between enqueuing the multiplications and the second bingo variable read.
\Cref{fig:spurious-stalls} shows the effect of this:
With 22 multiplications, followed by 0 to 30 \texttt{pxor} instructions, the second bingo variable read is still executed immediately without stalling the backend.
However, with 23 multiplications, we observe that the second bingo variable read is delayed when we insert 5 or more \texttt{pxor} instructions.
With 24 multiplications, we observe the same delay when we insert 4 or more \texttt{pxor} instructions.
According to official AMD documentation~\cite{AMD2020OptimizationEPYC7003}, the frontend can dispatch up to 6 \uops to the FPU per cycle.
We conclude that, with 23 or 24 multiplications, there is a short time window of about one cycle, after which the Zen 3 backend spuriously stalls.

With nanoBench~\cite{Abel2020nanobench} we measure that a single \texttt{rdpru} instruction (with \texttt{ecx = 1}) requires 17 CPU cycles on our machine.
Apparently, with 23 or 24 multiplications, the backend spuriously stalls before \texttt{rdpru} has finished reading the \texttt{APERF} counter, explaining the 2 \uops difference between earlier and our measurement methods.
We also run a similar experiment on our Zen 2 machine, where we fill the scheduler queue to its capacity limit of 16 \uops and insert up to 100 \texttt{pxor} instructions.
On that machine, we do not observe these spurious stalls, showing that these are specific to the Zen 3 microarchitecture.

\section{Scheduler Queue Usage of Various Instructions on Zen 3 and 4}\label{app:queue-instructions}
\Cref{tab:queue-instructions} shows the scheduler queue usage for various instructions on both the Zen 3 and the Zen 4 microarchitectures.
\begin{table}[ht]
  \caption{Scheduler Queue Usage of Various Instructions on Zen 3 and 4\textsuperscript{1}}
  \label{tab:queue-instructions}
  \resizebox{\hsize}{!}{
    \begin{tabular}{l ll l l}
      \toprule
        \textbf{Instruction} & \multicolumn{2}{l}{\textbf{Measurements}}                          & \textbf{Documentation} & \textbf{Comment}                                 \\
                                         & \parbox[l]{2.4cm}{Capacity}         & Scheduler        & ALU / Scheduler        &                                                  \\
      \midrule
        \texttt{idiv r10}                & 23                                  & Integer 0        & ALU0 / Integer 0       & rax=1, rdx=0, r10=1                              \\
        \texttt{movd xmm1, eax}          & 23                                  & Integer 0        & ALU2 / Integer 2       &                                                  \\
        \texttt{vmovd xmm1, rax}         & 23                                  & Integer 0        & ALU2 / Integer 2       &                                                  \\
        \texttt{cvtsi2sd xmm1, rax}      & 23$_3$ 22$_4$                       & Integer 0        & ALU2+FPU2/3 / Integer 2+FP0/1  &                                          \\
        \texttt{imul rax, 3}             & 24                                  & Integer 1        & ALU1 / Scheduler 1     &                                                  \\
        \texttt{stosb}                   & 22                                  & Integer 2        & undocumented (ucode)   &                                                  \\
        \texttt{lodsb}                   & 22                                  & Integer 3        & undocumented (ucode)   & \emph{delayable} due to partial register write   \\
        \texttt{lodsw}                   & 22                                  & Integer 3        & undocumented (ucode)   & \emph{delayable} due to partial register write   \\
        \texttt{lodsd}                   & 22                                  & Integer 3        & undocumented (ucode)   & \emph{delayable} due to false input dependency   \\
        \texttt{lodsq}                   & 22                                  & Integer 3        & undocumented (ucode)   & \emph{delayable} due to false input dependency   \\
        \texttt{bsf rbx, rax} \qquad(Zen 3) & \phantom{0}7                     & Integer 3        & undocumented           & not \emph{single-\uop}                           \\
        \texttt{bsf rbx, rax} \qquad(Zen 4) & 89                               & Integer 0/1/2/3  & ALU0/1/2/3 / Integer 0/1/2/3 & not \emph{targeted}                        \\
        \texttt{bsr rbx, rax} \qquad(Zen 3) & \phantom{0}7                     & Integer 3        & undocumented           & not \emph{single-\uop}                           \\
        \texttt{bsr rbx, rax} \qquad(Zen 4) & 89                               & Integer 0/1/2/3  & ALU0/1/2/3 / Integer 0/1/2/3 & not \emph{targeted}                        \\
        \texttt{rol rax, 3}              & 46                                  & Integer 1/2      & ALU1/2 / Integer 1/2   & not \emph{targeted}                              \\
        \texttt{shr rax, 3}              & 46                                  & Integer 1/2      & ALU1/2 / Integer 1/2   & not \emph{targeted}                              \\
        \texttt{add rax, 3}              & 91$_3$ 89$_4$                       & Integer 0/1/2/3  & ALU0/1/2/3 / Integer 0/1/2/3 & not \emph{targeted}                        \\
        \texttt{vaddsd xmm0, xmm0, xmm0} & 127$_3$ 124$_4$                     & FP0/1            & FPU0/1/2/3 / FP0/1     & not \emph{targeted}                              \\
        \texttt{divsd xmm0, xmm0}        & 127$_3$ 124$_4$                     & FP0/1            & FPU1 / FP1             & not \emph{targeted}                              \\
        \texttt{sqrtsd xmm0, xmm0}       & 127$_3$ 124$_4$                     & FP0/1            & FPU1 / FP1             & not \emph{targeted}                              \\
        \texttt{xor rax, rax}            & -                                                           & none             & none (zeroing idiom)   & not \emph{targeted}                              \\
        \texttt{mov rbx, rax}            & -                                                           & none             & none (rename only)     & not \emph{targeted}                              \\
      \bottomrule
    \end{tabular}\\\,\\
  }
\footnotesize{\textsuperscript{1}\,We see small differences in the capacity measurements for most instructions, the Zen version is denoted as a subscript in these cases.
The reason for that is not a different scheduler queue size but smaller differences in the microarchitecture, like different scheduling of other instructions executed for the measurement.
We only observe a large difference between Zen 3 and Zen 4 with the \texttt{bsf} and \texttt{bsr} instruction.
While consisting of multiple \uops{} executed only on ALU3 in Zen 3, they are single \uop{} instructions that can be executed on all integer ALUs in Zen 4.}
\end{table}

\section{Source Code Excerpts for the JavaScript-based Covert Channel}\label{app:javasquip}
The following listings show the attack code for our JavaScript-based Covert Channel and a disassembly of the JIT-compiled code generated by Firefox:
\begin{lstlisting}[frame=,language=JavaScript,numbers=left,caption={The JavaScript-based covert channel attack code.},label={lst:javasquip}]
// Shared array buffer
var timer = new Uint32Array(...);

function squip(start_value, factor) {
  start_value = +start_value;
  factor = factor|0;

  let dummy = +start_value;
  let dummy2 = start_value|0;

  let bingo1 = Atomics.load(timer, 0)|0;

  dummy = +Math.sqrt(+dummy); // repeat 12 times

  dummy = dummy|0;

  dummy = Math.imul(dummy|0, factor|0)|0; // repeat 20 times

  let bingo2 = Atomics.load(timer, 0)|0;

  return {
    dummy: dummy|0,
    time: ((bingo1|0) != (bingo2|0))|0
  };
}
\end{lstlisting}
\begin{lstlisting}[frame=,language={[x64]Assembler},numbers=left,caption={Disassembly of the JIT-compiled code generated by Firefox.},label={lst:javasquip-disasm}]
mov ebx, [rsi+rbp*4]

; ...

xorpd xmm1, xmm1
cvtsi2sd xmm1, eax
.rept 12
sqrtsd xmm1, xmm1
.endr
cvttsd2si rdi, xmm1

cmp rdi, 1
jo 570h
mov edi, edi
mov r11, 0fffe000000000000h
xor r11, [r9 + 58h]
mov r9, r11
shr r11, 2fh
jnz 69dh
mov r8, 22aa1cc9c660h
cmp r9, r8
jnz 6a7h

.rept 20
imul edi, ecx
.endr
mov r8d, [rsi+rbp*4]
\end{lstlisting}
Lines 25 to 27 show the \texttt{imul} instructions for priming Scheduler 1, which depend on the \texttt{sqrtsd} instructions and the subsequent integer conversion in lines 7 to 10.
The bingo variable is read in lines 1 and 27, with the second read directly following the last \texttt{imul} instruction.

Firefox inserts some additional code (lines 12 to 22) between the integer conversion and the \texttt{imul} instructions, which does not hinder our measurements:
Lines 12, 14, 15, 17, 18, 20 and 21 are low-latency operations that only occupy the scheduler queues for a very short time.
Line 16 has a memory operand, however this operand is cache-hot, therefore it also has a low latency and does not prevent the second bingo variable read from being executed earlier in the non-contended case (see~\cref{sub:bingo-limits}).
Lines 13, 19 and 22 are jumps that are never taken, hence they are speculated over, after a few warm-up iterations.

\end{document}